\documentclass[nohyper,11pt,a4paper]{JHEP3}
\usepackage{epsfig}
\usepackage{psfrag}
\usepackage{graphicx}

\title{Cosmological Models and Renormalization Group Flow}

\author{Kristj\'an R.\ Kristj\'ansson and L\'arus Thorlacius\\ 
University of Iceland, Science Institute \\
Dunhaga 3, 107 Reykjavik, Iceland\\
E-mail: \email{kristk@hi.is}, \email{lth@hi.is}}

\abstract{We study cosmological solutions of Einstein gravity with a 
positive cosmological constant in diverse 
dimensions.  These include big-bang models that re-collapse, 
big-bang models that approach de~Sitter acceleration at late times, 
and bounce models that are both past and future asymptotically 
de~Sitter.  The re-collapsing and the bounce geometries are all 
tall in the sense that entire spatial slices become visible to a 
comoving observer before the end of conformal time, while the 
accelerating big-bang geometries can be either short or tall.  
We consider the interpretation of these cosmological solutions as 
renormalization group flows in a dual field theory and give a 
geometric interpretation of the associated c-function as the area 
of the apparent cosmological horizon in Planck units.  The covariant 
entropy bound requires quantum effects to modify the early causal 
structure of some of our classical big-bang solutions.}


\preprint{RH-05-2002}

\begin{document}




\section{\bf Introduction}

There has been considerable recent interest in the study of gravity with 
a positive cosmological constant $\Lambda$.  This is in part due to 
cosmological observations indicating that our universe is undergoing
accelerated expansion compatible with a small positive 
$\Lambda$ \cite{supernov}, but also because of various interesting 
theoretical issues that arise, 
see {\it e.g.} \cite{hull} - \cite{strominger}.  
In particular, it has been conjectured that gravity in $n{+}1$-dimensional 
de~Sitter spacetime has a dual description as a Euclidean conformal field 
theory in $n$ dimensions \cite{strominger}.  
The proposed dS/cft duality has been extended to more general spacetime 
geometries that are asymptotically de~Sitter, with cosmological evolution 
interpreted as a renormalization group flow on the field theory 
side \cite{strominger2,vijay}. 
The physics is in some respects analogous to that of the much better
understood adS/cft duality and related renormalization group 
flows \cite{adsreview}, but other aspects are clearly quite different.
For one thing, supersymmetry is absent from de~Sitter gravity.  
Another important feature is the finite area of the de~Sitter event 
horizon \cite{gibhaw}, which has been argued to put a finite upper limit 
on the number of available degrees of freedom in de~Sitter gravity
\cite{banks,bousso,fischler}.  This would in turn imply that the proposed
dS/cft duality could only be exact in the limit of infinite de~Sitter 
entropy, that is vanishing $\Lambda$ \cite{dls}.  

In the present paper we study dS/cft related issues in a relatively
simple context.  In section~\ref{models} we present a number of analytic 
cosmological solutions of gravity with positive $\Lambda$ in various 
spacetime dimensions.  We restrict our attention to homogeneous and 
isotropic cosmological models with perfect fluid matter and linear 
equations of state.  This rather 
restricted framework yields a surprisingly rich set of exact solutions, 
which include big-bang models that approach de~Sitter acceleration in the
asymptotic future, re-collapsing big-bang models, and bounce geometries
that approach de~Sitter behavior both in the asymptotic past and 
asymptotic future.  In section~\ref{confdiagrams} we analyze the causal 
structure of the various solutions.  This is followed, in 
section~\ref{horizons}, by a discussion of cosmological horizons, both 
event horizons and apparent horizons.  In section~\ref{cfunct} we 
consider the interpretation of cosmological evolution as a 
renormalization group flow in a dual field theory.  In particular, 
we identify the associated c-funtion with the area of the apparent 
cosmological horizon in Planck units.  The c-theorem then becomes 
a geometric statement about the increase of the apparent horizon area
in an expanding asymptotically de~Sitter spacetime. We close with the
observation that the covariant entropy bound \cite{boussoreview} is
violated at very early times in some of these models.  The violation
may, however, be traced to a region of spacetime where quantum gravity
effects are expected to be important. 

During the course of this work, there has been considerable activity
in this area of research and papers have appeared on related topics.  
We note in particular \cite{lmm}, which studies cosmological models in 
Einstein gravity coupled to a scalar field with emphasis on their 
interpretation as renormalization group flows, and \cite{medved}, 
which considers perfect fluid models and contains some of the solutions 
that we present below. 

\section{\bf Cosmological models with positive vacuum energy}
\label{models}

We consider $n{+}1$-dimensional homogeneous and isotropic 
cosmological models with $n\geq 2$ and a positive cosmological constant. 
The analysis of such models in 3+1 dimensions is standard material, see for
example \cite{mtw, hawell}, and many qualitative features carry over to 
general dimensions. The metric can be put into Robertson-Walker form,
\begin{equation}
ds^2 = - dt^2 + R^2(t)\left(
{dr^2\over 1-kr^2} + r^2 d\Omega_{n-1}^2\right) \,,
\label{rwmetric}
\end{equation}
where $d\Omega_{n-1}^2$ is the line element of an $n{-}1$-dimensional unit
sphere.  The geometry of constant $t$ slices depends on the sign of the 
parameter $k$, being spherical for $k>0$, flat for $k=0$, and hyperbolic 
for $k<0$.  It is common to rescale the spatial coordinates in such a 
way as to make $k$ equal to $1,0,$ or $-1$, but we will not do that here.
Instead we choose to retain the freedom to rescale the coordinates and 
our results will be given in terms of parameters that are invariant under 
such rescalings.
 
We take the matter to be a perfect fluid.  Einstein's equation and 
energy-momentum conservation can then be expressed as
\begin{eqnarray}
\left({dR\over dt}\right)^2 & = & 
{2\over n(n-1)}\left[\Lambda+8\pi G\rho\right]R^2 -k  \,, 
\label{einstein} \\
{d\rho\over dR} & = &
-{n\over R}(\rho + P) \,. 
\label{eoms}
\end{eqnarray}
We will restrict our attention to equations of state of the form 
\begin{equation}
P = \alpha \rho \, ,
\label{eqstate}
\end{equation}
with constant $\alpha$.
Two cases of interest are radiation (with $\alpha=1/n$) 
and dust (with $\alpha=0$).

Equation (\ref{eoms}) integrates to 
\begin{equation}
\log{\rho}=-(1+\alpha)n\, \log{R} + const \,.
\end{equation}
The integration constant can be expressed in terms of the 
scale factor at `cross-over', {\it i.e.} when the matter energy density 
equals the vacuum energy density,
\begin{equation}
R = R_* \longleftrightarrow
\rho = \rho_\Lambda = {\Lambda\over 8\pi G} \,,
\end{equation}
or equivalently
\begin{equation}
{\rho \over \rho_\Lambda}=\left({R_*\over R}\right)^{(1+\alpha)n} \,.
\label{rhorad}
\end{equation}
Inserting this into equation (\ref{einstein}) gives
\begin{equation}
\left({dR\over dt}\right)^2 = {2\Lambda\over n(n-1)}
\left[1+\left({R_*\over R}\right)^{(1+\alpha)n}\right]R^2-k \,.
\end{equation}
Now introduce a dimensionless time variable 
$\tau=\sqrt{2\Lambda\over n(n-1)}\,t$ 
and define $u=R/R_*$ to obtain
\begin{equation}
\left({du\over d\tau}\right)^2 = u^2+ u^{-(1+\alpha)n+2}-\kappa \,,
\label{dusquared}
\end{equation}
where $\kappa = {n(n-1)\,k \over 2\Lambda R_*^2}$ 
is a dimensionless quantity.  We observe that, while both $k$ and $R_*$ 
change under rescaling of the spatial coordinates, $\kappa$ is invariant.
We also note that equation (\ref{dusquared}) only involves the 
combination $(1+\alpha)n$, which means in particular that the
evolution for pressureless dust in $n{+}1$ dimensions is identical 
to that for radiation in one less dimension.

\FIGURE{
    \psfrag{u}{\footnotesize{$u$}}
    \psfrag{u0}{\footnotesize{$u_0$}}
    \psfrag{-k0 }{\footnotesize{$-\kappa_0$}}
    \psfrag{v(u)}{\footnotesize{$v(u)$}}
    \epsfig{file=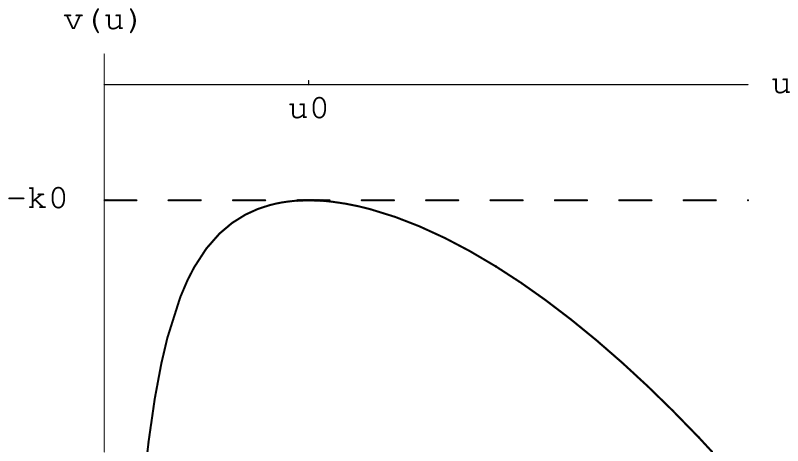, width=6.0cm}
    \caption{A typical potential for the equivalent particle motion.}
    \label{effpot} 
}

Various values of $n$ and $\alpha$ yield simple analytic solutions
and we analyze some of these below.  We can, however, get a 
qualitative picture of the solutions for any $n$ and $\alpha$ by 
rewriting equation (\ref{dusquared}) as 
\begin{equation}
\left({du\over d\tau}\right)^2 + v(u) = -\kappa \,,
\end{equation}
and proceeding by analogy with one-dimensional particle motion in
the potential
\begin{equation}
v(u) = -u^2-u^{-(1+\alpha)n+2} \,.
\label{effpotential}
\end{equation}
The corresponding analysis for 3+1-dimensional cosmological models
is carried out in \cite{mtw}.
A typical potential is drawn in figure~\ref{effpot}.  The parameter
$-\kappa$ plays the role of the total energy of the particle.  For
$\kappa$ below a certain value $\kappa_0$, or equivalently, a total
energy above the maximum of the potential, the particle can roam 
from $u=0$ to $u\rightarrow\infty$ and one obtains big-bang 
solutions that eventually approach de~Sitter acceleration.
Time-reversed solutions, describing a past asymptotically de~Sitter
universe that ends in a big crunch singularity, are also allowed.

Geometries with $\kappa>\kappa_0$, on the other hand, correspond
to a total energy below the maximum of $v(u)$.  The particle is 
then either confined to small values of $u$, resulting in big-bang
solutions that re-collapse to a big crunch, or the particle comes in 
from large $u$ and is reflected off the potential barrier.  These 
reflecting `bounce' solutions are less realistic than the 
big-bang geometries from the point of view of cosmology but, being
both past and future asymptotically de~Sitter, they may prove useful
for exploring the dS/cft correspondence.  Finally, we have the 
Einstein static universe with $\kappa=\kappa_0$, corresponding to 
the particle perched in unstable equilibrium at the top of the 
potential.  

\subsection{Spatially flat models}
\label{flatmod}

It is always useful to have explicit solutions to work with and in
the following we present several examples.  We begin by restricting
our attention to spatially flat models with $\kappa=0$.  In this
case, an exact solution to the evolution equation (\ref{dusquared})
for any combination of $n$ and $\alpha$ is given by \cite{medved}
\begin{equation}
\label{zerokappa}
u(\tau)=\sinh^\beta ({\tau\over \beta}) \,,
\end{equation}
with $\beta={2\over (1+\alpha)n}$.  These solutions describe big-bang
geometries that expand from an initial singularity.  They eventually 
enter an accelerating phase and are future asymptotically de~Sitter.  
We have placed the origin of our time coordinate, $\tau=0$, at the initial 
singularity.  For small $\tau$ the scale factor is
\begin{equation}
u(\tau)\approx \left({\tau\over \beta}\right)^\beta \,,
\end{equation}
which is the expected behavior for the early universe.  
At late times the cosmological term takes over and we instead find
\begin{equation}
u(\tau)\approx 2^{-\beta} \, e^\tau \,,
\end{equation}
which is de~Sitter behavior.

These are by no means generic models, given that the matter energy 
density has been fine tuned to give $k=0$, but the special case
$\beta=2/3$, which is found in \cite{peebles} and corresponds to 
pressureless dust in 3{+}1 dimensions, appears to fit the observed 
universe rather well today \cite{supernov}.
In order to exhibit the full range of behavior outlined earlier, including
re-collapsing universes and bounce geometries, one has to allow 
generic matter energy density, {\it i.e.} $\kappa\neq 0$.  In this 
more general case, explicit solutions are only found for certain 
combinations of $n$ and $\alpha$ and we work out some of these below.

\subsection{Radiation models in 3+1 dimensions}

Let us consider $n=3$ and matter in the form of radiation with
$\alpha={1/3}$.  The same evolution is obtained for $n=4$ 
and $\alpha=0$, which corresponds to dust in $4+1$ dimensions.  
The equivalent one-dimensional potential reduces to $v(u)=-u^2-u^{-2}$, 
with a maximum at $u_0=1$ that corresponds to $\kappa_0=2$.  It is then 
straightforward to integrate equation (\ref{dusquared}) and obtain
the scale factor in closed form. The qualitative physical behavior of 
these solutions is well known \cite{mtw} and depends on the value
of $\kappa$ and the boundary conditions imposed on the integration.

\subsubsection{Big-bang solutions}
We begin with big-bang geometries that expand from an initial 
singularity.  The solution takes the form
\begin{equation}
u(\tau) = \sqrt{\sinh{2\tau}+(\kappa/2)(1-\cosh{2\tau})}\,,
\label{radscale}
\end{equation}
where we have used time-translation symmetry to put the initial 
singularity at $\tau=0$.  This relatively simple expression for the
scale factor is valid for any value of $\kappa$ and thus covers all 
three cases of positive, flat, and negative spatial curvature.  
All these models have the same initial rate of expansion, 
\begin{equation}
u(\tau) \approx \sqrt{2\tau} \,,
\end{equation}
as expected for a radiation-dominated universe in $3+1$ dimensions (or 
pressureless dust in $4{+}1$ dimensions) but the late time behavior is 
governed by the value of $\kappa$.  For matter energy 
densities, such that $\kappa<2$, the vacuum energy 
density eventually dominates and the scale factor ultimately approaches 
the exponential expansion of de Sitter spacetime,
\begin{equation}
u(\tau) \approx  {\sqrt{2-\kappa}\over 2}\, e^\tau 
\quad {\rm as} \quad \tau\rightarrow\infty\,.
\end{equation}
Note that the dividing line between eternal expansion
and re-collapse does not occur at the spatially flat solution.  In the
presence of a cosmological constant we can in fact have an ever expanding
geometry with spherical spatial sections that are finite in extent at any
given cosmic time.  In such a universe the matter energy density is
large enough to give closed spatial sections but not enough to overcome
the effect of the positive cosmological constant.

If, on the other hand, the matter energy density is large enough
to give $\kappa>2$ the universe expands to a maximum size,
\begin{equation}
u_{max} = \sqrt{{\kappa\over 2}-\sqrt{{\kappa^2\over 4}-1}} \,,
\end{equation}
and then re-contracts back to zero scale factor at a finite time,
\begin{equation}
\tau_f = {1\over 2}\,\log\left({\kappa+2\over\kappa-2}\right) \,.
\end{equation}
As $\kappa\rightarrow 2^+$ the total lifetime, $\tau_f$, before the 
universe 
ends in a big-crunch, becomes arbitrarily long.  
As $\kappa\rightarrow\infty$, however, the matter energy dominates
and the total lifetime is vanishing on the timescale set by $\Lambda$.  
One can then expand the hyperbolic functions in equation 
(\ref{radscale}) to leading orders to recover the behavior of a 
re-collapsing $3{+}1$-dimensional universe with radiation and $\Lambda=0$,
\begin{equation}
u(\tau)\propto 
\sqrt{{\tau\over\tau_f}-{\tau^2\over\tau^2_f}} \,.
\end{equation}

The ratio of matter and vacuum energy densities is given by
\begin{equation}
{\rho \over \rho_\Lambda} = {1\over u^4(\tau)} \,.
\label{uifjorda}
\end{equation}
Since $u_{max}<1$ for all $\kappa>2$, the matter energy density 
exceeds the vacuum energy density at all times in the re-collapsing 
geometries. 

\FIGURE{
    \psfrag{u}{$u$}
    \psfrag{tau}{$\tau$}
    \epsfig{file=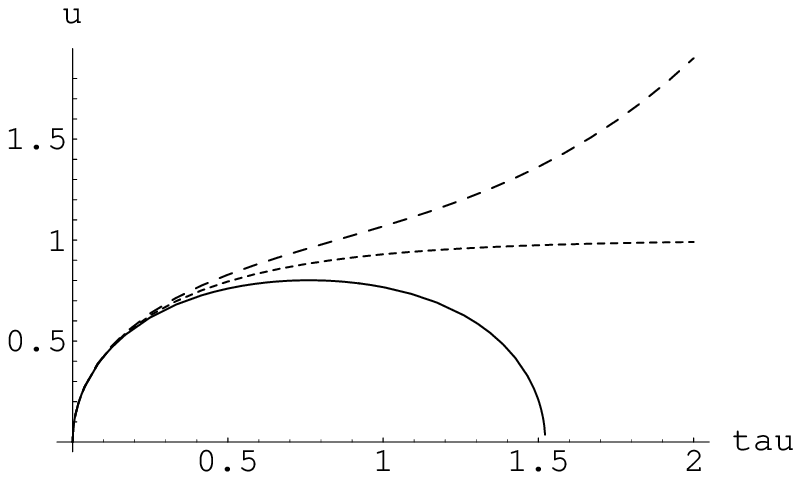, width=6.5cm}
    \caption{The scale factor $u$ as a function of $\tau$ for big-bang 
    models with matter in form of radiation and
    $\kappa = 2.2$ (solid curve),
    $\kappa = 2.0$ (dotted curve), and $\kappa = 1.8$ (dashed curve).}
    \label{scalebb} 
}

Finally, there is the borderline case of $\kappa=2$ where the expansion 
slows down with time and the scale factor approaches a fixed value from 
below, 
\begin{eqnarray}
u(\tau)&=& \sqrt{1-e^{-2\tau}} \nonumber \\
&\rightarrow&  1 \quad {\rm as} \quad \tau\rightarrow\infty \,.
\end{eqnarray}
This solution, which approaches the Einstein static universe, 
is unstable in the sense that if $\kappa$ deviates at all
from $2$ the universe eventually finds itself collapsing back
to zero size or undergoing exponential expansion. 
Figure~\ref{scalebb} depicts the 
scale factor for big-bang models with different values of $\kappa$.

\subsubsection{Bounce solutions}

These solutions only occur for $\kappa>2$ and therefore have spherical 
spatial sections.  The spacetime geometry is non-singular and 
asymptotically de Sitter both in the past and future.   Initially the 
scale factor is decreasing but 
after reaching a finite minimum value it bounces back and eventually 
approaches exponential de~Sitter expansion.  The `bounce' solutions 
are given by 
\begin{equation}
\label{bounce}
u(\tau) = \sqrt{{\kappa\over 2} 
+ \sqrt{{\kappa^2\over 4}-1}\,\cosh 2\tau} \,,
\end{equation}
where we have chosen the zero of the time coordinate to be when the scale 
factor takes its minimum value,
\begin{equation}
u_{min} = u(0) = \sqrt{{\kappa\over 2}
+ \sqrt{{\kappa^2\over 4}-1}} \,.
\end{equation}  
Note that $u_{min}>1$ for all $\kappa>2$ so, by equation 
(\ref{uifjorda}), the vacuum energy exceeds the matter energy at all 
times.  Examples of bounce geometries are displayed in 
figure~\ref{scalebounce}.  
\FIGURE{
    \psfrag{u}{$u$}
    \psfrag{tau}{$\tau$}    
    \epsfig{file=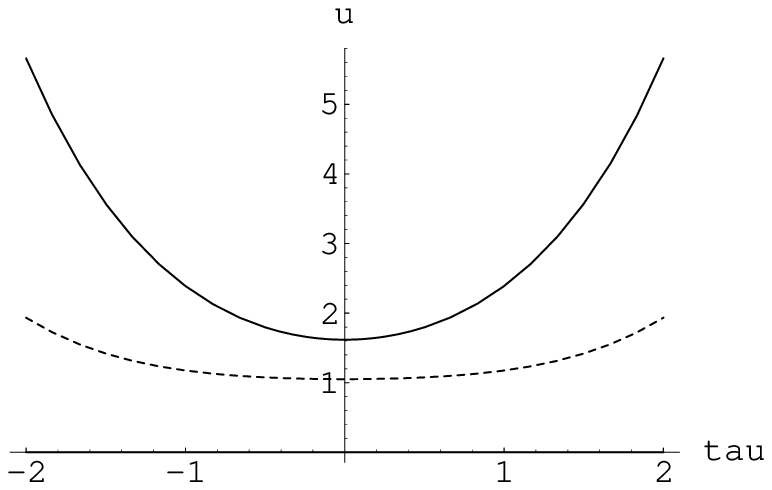, width=5cm}
    \caption{The scale factor $u$ as a function of $\tau$ 
        for bounce models in 3+1 dimensions.
        The dotted curve is for $\kappa =2.01$ and the 
        solid curve is for $\kappa = 3$. }
    \label{scalebounce} 
}

In the limit $\kappa\rightarrow 2^+$ the solution spends a long time 
`near' the Einstein static geometry.  In the opposite limit, 
$\kappa\rightarrow\infty$, it instead approaches pure 
de~Sitter space.  To see this, consider the energy density at 
$\kappa\gg 1$, for which $u_{min}\approx\sqrt{\kappa}\gg 1$.  It then
follows from equation (\ref{uifjorda}) that the matter energy
density is vanishing compared to the vacuum energy at all times
and the geometry must reduce to the de~Sitter vacuum.
The same conclusion can also be reached by explicit calculation, 
writing the Robertson-Walker line element (\ref{rwmetric}) in terms 
of our dimensionless variables, inserting (\ref{bounce}) for $u(\tau)$,
and observing that the line element reduces in the 
$\kappa\rightarrow\infty$ limit to that of de~Sitter spacetime in 
global coordinates.

All the $\kappa>2$ solutions are time-symmetric around some reference time, 
at which the scale factor is at an extremum.   The $\kappa<2$ big-bang
solutions are, on the other hand, asymmetric between the initial 
singularity and the future asymptotic de~Sitter expansion, but 
time-reversed solutions, with de~Sitter contraction in the asymptotic past 
and ending in a big crunch singularity, are also allowed. 

\subsection{Dust in 3+1 dimensions or radiation in 2+1 dimensions}

Qualitative features of $3{+}1$-dimensional cosmological models, with 
$\Lambda>0$ and matter in the form of dust, are outlined in \cite{mtw,hawell}.
We obtain such models by setting $n=3$ and $\alpha=0$ leading to an
effective one-dimensional potential $v(u)=-u^2-u^{-1}$.  The same 
potential is obtained for a 2+1-dimensional radiation-filled universe
with $n=2$ and $\alpha=1/2$.  The potential has its maximum at 
$u=u_0=2^{-1/3}$ and its value at the maximum gives 
$\kappa_0=3\cdot 2^{-2/3}$.  

The analytic solution for $\kappa=0$ is obtained by setting 
$\beta=2/3$ in equation (\ref{zerokappa}) but numerical evaluation is 
required for general values of $\kappa$.  The solutions include 
big-bang models that either approach de~Sitter acceleration or 
re-collapse, depending on the value of $\kappa$ relative to $\kappa_0$.
There are also bounce solutions for $\kappa>\kappa_0$, and a dust-filled
static universe (or radiation-filled universe in 2+1 dimensions) with
$\kappa=\kappa_0$ and $u=u_0$.

\subsection{Dust models in 2+1 dimensions}
\label{twoddust}

Another class of exact solutions is obtained for
pressureless dust in 2+1 dimensions, with $n=2$ and $\alpha=0$.  
They do not correspond to realistic cosmological models but being 
asymptotic to $2{+}1$-dimensional de~Sitter space they provide a
particularly simple context in which to explore dS/cft ideas.
The equivalent one-dimensional potential is $v(u)=-u^2-1$ and is
drawn in figure~\ref{dustpot}.  Note that the maximum of the potential now
occurs at $u=0$.

\FIGURE{
    \psfrag{u}{\footnotesize{$u$}}
    \psfrag{-1}{\footnotesize{$\!\!\!\!-1$}}
    \psfrag{v}{\footnotesize{$v(u)$}}   
    \epsfig{file=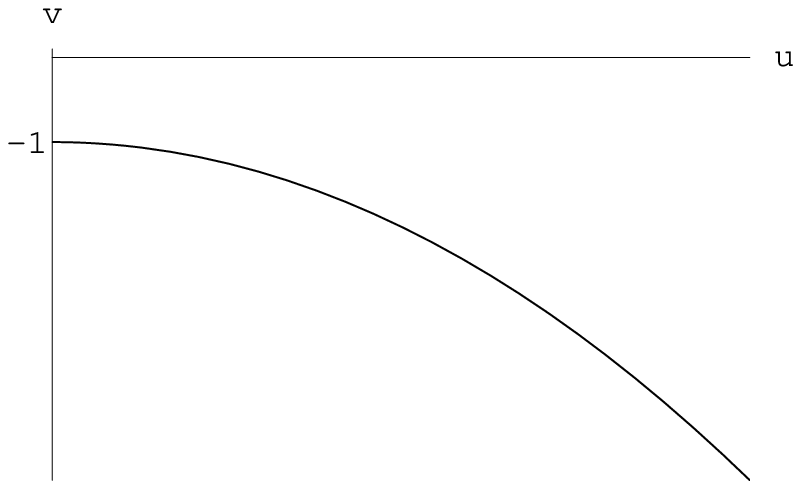, width=5cm}
    \caption{The equivalent potential for  
        for dust models in 2+1 dimensions.}
    \label{dustpot} 
}

For $\kappa<1$, one finds a family of big-bang solutions that eventually
approach de~Sitter expansion,
\begin{equation}
u(\tau) = \sqrt{1-\kappa}\,\sinh\tau \,.
\label{twodbb}
\end{equation}
There are of course also the corresponding time-reversed big-crunch
solutions.

For $\kappa>1$, there is a minimum allowed value for $u$ and big-bang
solutions which expand from $u=0$ are ruled out.  There are, however,
bounce solutions of the form
\begin{equation}
u(\tau) = \sqrt{\kappa-1}\,\cosh\tau \,,
\label{twodbounce}
\end{equation}
which, like their $3{+}1$-dimensional counterparts, approach pure de~Sitter
space in global coordinates in the $\kappa\rightarrow\infty$ limit.

Finally, we have the dividing value $\kappa=1$ for which the solution
takes the form
\begin{equation}
u(\tau) = u_0\,e^{\pm\tau} \,,
\end{equation}
with the sign in the exponent depending on whether the geometry is
expanding or contracting with time.  We note that, with matter in the
form of pressureless dust, there is no analog of the Einstein static
universe in our $2{+}1$-dimensional cosmology and no re-collapsing big-bang
geometries.

\subsection{Negative pressure matter}
\label{negpress}

Let us finally consider models with $\alpha<0$, where the matter has 
negative pressure.  Such equations of state occur for various
dynamical matter systems, for example a minimally coupled scalar field.
Negative pressure matter has been invoked to explain the observed cosmic
acceleration in the absence of a cosmological constant \cite{quint}.
Our focus here is on gravity in asymptotically de~Sitter spacetime so 
we retain the cosmological constant term and take the negative pressure 
fluid to have $-1<\alpha<0$.  We note that this form of matter satisfies
the dominant energy condition,
\begin{equation}
\rho\geq \vert P\vert \,.
\end{equation}
For equations of state in the range $-1+\frac{2}{n}<\alpha<0$ the 
equivalent particle potential (\ref{effpotential}) has a form as shown
in figure~\ref{effpot} and we obtain the corresponding types of 
cosmological solutions, including ever-expanding big-bang models 
for $\kappa<\kappa_0$, and re-collapsing big-bang models and bounces
for $\kappa>\kappa_0$.  At $\alpha=-1+\frac{2}{n}$ we get the same 
solutions as in the $2{+}1$-dimensional dust models described above.
In the remaining range $-1<\alpha<-1+\frac{2}{n}$ the particle potential
approaches $v=0$ as $u\rightarrow 0$ but the rate of approach depends
on the value of $\alpha$ relative to $-1+\frac{1}{n}$, as illustrated in 
figure~\ref{potrate}.
\FIGURE{
    \psfrag{u}{\footnotesize{$u$}}
    \psfrag{vu}{\footnotesize{$v(u)$}}   
    \epsfig{file=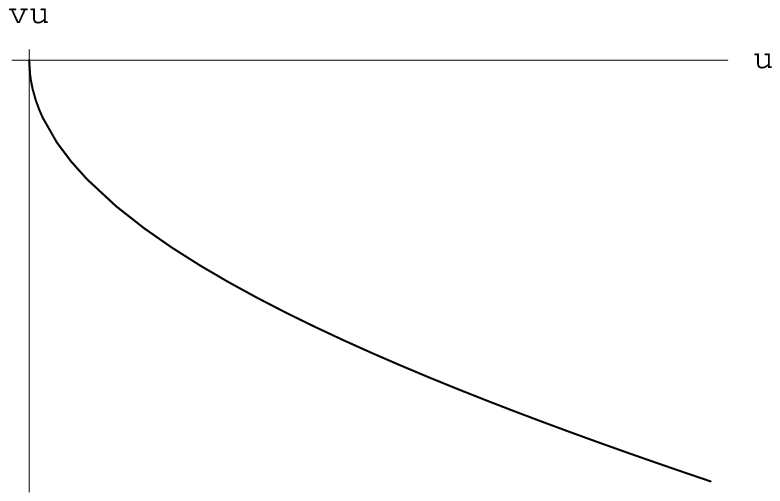, width=3.8cm}\hfill
    \epsfig{file=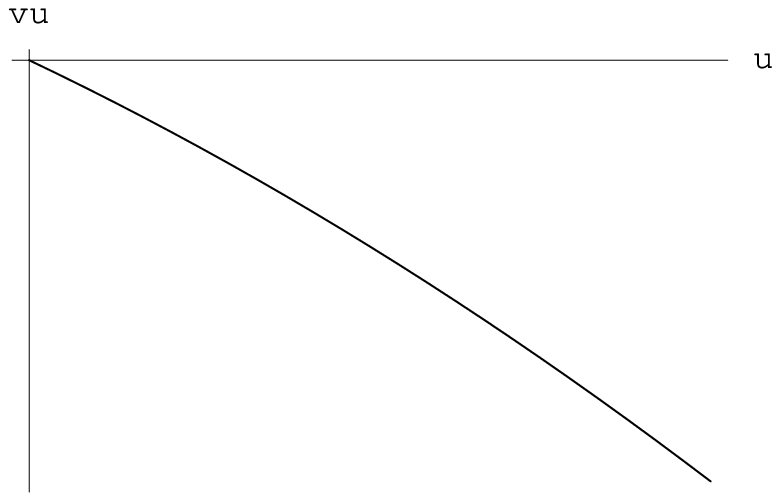, width=3.8cm}\hfill
    \epsfig{file=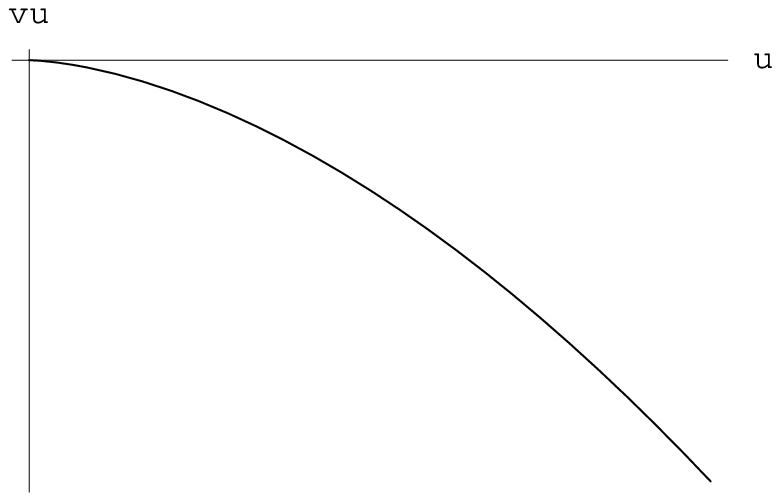, width=3.8cm}
    \caption{The particle potential close to $u=0$ for three
      different equations of state. 
      (a) For $-1 + {1 \over n} < \alpha < -1 + \frac{2}{n}$ the slope of the 
      potential approaches $-\infty$ as $u \rightarrow 0$,
      (b) for $\alpha = -1 + \frac{1}{n}$ the slope is $-1$ and  
      (c) for $-1 < \alpha <-1 + \frac{1}{n}$ the slope is zero at $u=0$.}
    \label{potrate} 
}
In all three cases, the solutions with $\kappa<0$ are big-bang models that
accelerate forever, while only bounce solutions are obtained for
$\kappa>0$.  Spatially flat solutions may be obtained as 
$\kappa\rightarrow 0$ limits of either type of model but, as we will
see later on, the two limits do not give the same geometry.  

We can obtain exact solutions for $\alpha=-1+\frac{1}{n}$ which is 
the case shown in figure~\ref{potrate}b.  Then the particle potential 
reduces to $v(u)=-u^2-u$. 
For $\kappa\leq 0$ one finds a family of big-bang solutions,
\begin{equation}
u(\tau) = \sinh^2 \frac{\tau}{2} + \sqrt{-\kappa}\sinh \tau \, ,
\label{npbb}
\end{equation}
that approach de~Sitter expansion at late times.  The time-reversed 
big-crunch solutions are also allowed.

For $\kappa > 0$, on the other hand, there are only bounce solutions
with
\begin{equation}
u(\tau) = \sqrt{\kappa + \frac{1}{4}}\cosh \tau -\frac{1}{2}\, .
\label{npbounce}
\end{equation}
Note that for $\kappa = 0$, the big-bang solution may be extended to 
cover negative $\tau$ as well, and then it formally agrees with the 
$\kappa=0$ bounce solutions.  In section~\ref{nppenr} we will see, 
however, that the $\kappa\rightarrow 0$ bounce geometry has vanishing 
spatial volume at the symmetry point $\tau=0$, while the 
$\kappa=0$ big-bang geometries (\ref{zerokappa}) have infinite spatial 
sections at all finite $\tau>0$.

\section{Penrose diagrams}
\label{confdiagrams}

Now consider the global causal structure of the various cosmological 
solutions that we have presented.  For a Robertson-Walker metric 
(\ref{rwmetric}), conformal time $\eta$ is defined through
\begin{equation}
d\eta={dt\over R(t)}\,,
\label{conformaltime}
\end{equation}
so that  
\begin{equation}
\label{conformalmetric}
ds^2 = 
R^2(t(\eta)) \left(-d\eta^2+{dr^2\over 1-kr^2}+r^2d\Omega^2_{n-1}\right) \,.
\end{equation} 
The line element can be expressed in terms of dimensionless 
variables as follows,
\begin{equation}
\label{dless}
{2\Lambda\over n(n{-}1)}\, ds^2={u^2(\tau)\over \vert\kappa\vert}\,
 \left(-d\tilde{\eta}^2+d\chi^2+f(\chi)^2d\Omega^2_{n-1}\right) \,,
\end{equation}
with $d\tilde{\eta}=\vert k\vert^{1/2}d\eta
=\vert\kappa\vert^{1/2}d\tau/u(\tau)$ and
\begin{equation}
f(\chi)=\vert k\vert^{1/2}\,r=\left\{ \>\>
\begin{array}{ccc}
\sin{\chi} & \qquad {\rm if} \qquad & \kappa>0 \,, \cr
\chi & \qquad {\rm if} \qquad & \kappa=0 \,, \cr
\sinh{\chi} & \qquad {\rm if} \qquad & \kappa<0 \,. 
\end{array} \right.
\label{ffunc}
\end{equation}
For the spatially closed models with $\kappa>0$, the radial variable 
$\chi$ is the polar angle of an $n$-sphere and has a finite range 
$0\leq\chi\leq\pi$, while for $\kappa\leq 0$ we have $0\leq\chi<\infty$.

Penrose diagrams are plots of conformal time against the radial variable.  
For the geometries with $\kappa\neq 0$ it is convenient to use the 
dimensionless variables $\tilde{\eta}$ and $\chi$, but for $\kappa=0$
the dimensionless variables are degenerate and one has to use $\eta$ and
$r$ instead.  Each point in the diagram represents a transverse
$n{-}1$~sphere of proper `area' 
\begin{equation}
{\cal A} = 
a_{n-1}\left(\vert\kappa\vert^{-1/2}u(\tau) f(\chi)\right)^{n-1}
\left({n(n-1)\over 2\Lambda}\right)^{(n-1)/2} \,,
\label{properarea}
\end{equation}
where $a_{n-1}=2\pi^{n/2}/\Gamma({n\over 2})$ is the area of the 
$n{-}1$~dimensional unit sphere.  This expression for the transverse 
area has a smooth limit as $\kappa\rightarrow 0$ and may be used for all 
our solutions.  Radial null-curves appear in a Penrose diagram as straight 
lines at $45^\circ$ angle from the vertical.

\subsection{Penrose diagrams for spatially flat models}
We first construct Penrose diagrams for the spatially flat models 
of section~\ref{flatmod}.  Conformal time is obtained by inserting the 
scale factor (\ref{zerokappa}) into equation (\ref{conformaltime}) and 
integrating over time. These models all accelerate forever in comoving 
time but conformal time nevertheless remains finite as 
$\tau\rightarrow\infty$.  In some cases, however, the expansion starts 
off too slow for conformal time to converge early on.  

\FIGURE{
        \psfrag{a}{(a)}
        \psfrag{b}{(b)}
        \includegraphics{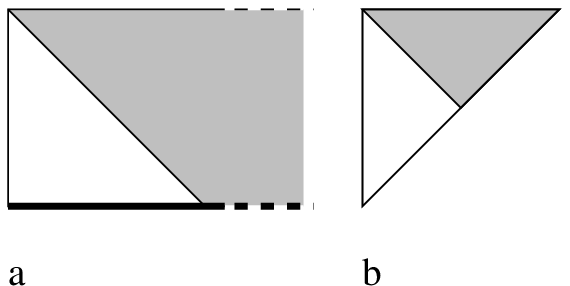}
        \caption{Penrose diagrams for spatially flat and hyperbolic 
        geometries: (a) with $\beta<1$, (b) with $\beta\geq 1$.}
        \label{pen_normal}
}

For $\beta<1$, {\it i.e.} $(1+\alpha)n>2$, the integration
converges at both ends and conformal time may be defined 
\begin{equation}
\label{spatflateta}
\eta(\tau)=\sqrt{n(n-1)\over 2\Lambda R_*^2} \int_0^\tau 
{dx\over \sinh^\beta ({x\over\beta})} \, ,
\end{equation}
where we have chosen to put $\eta=0$ at the initial singularity.
The integral can be expressed in terms of the incomplete Euler beta 
function if desired.  The maximal conformal time is finite and given by
\begin{equation}
\label{eulerb}
\eta_{max} =\sqrt{n(n-1)\over 2\Lambda R_*^2}\, 2^{\beta{-}1} \beta
\, B\left(1{-}\beta,{\beta\over 2}\right) \, ,
\end{equation}
where $B(a,b)$ is the usual Euler beta function.
The resulting Penrose diagram is shown in 
figure~\ref{pen_normal}a.  The geometry has infinitely large spatial 
slices but only a finite spatial region is in the causal past of any
given comoving observer.

For $\beta\geq 1$ the integration in (\ref{spatflateta}) diverges at the
lower end.  As a result we find it more convenient to take
$\tau\rightarrow\infty$ as our reference point, $\eta=0$, when 
defining conformal time,
\begin{equation}
\eta(\tau)=-\sqrt{n(n-1)\over 2\Lambda R_*^2} \int_\tau^\infty
{dx\over \sinh^\beta ({x\over\beta})}  \,.
\label{diveta}
\end{equation}
Now define null coordinates
\begin{equation}
x^\pm = {1\over \sqrt{2}}(\eta\pm r) \,,
\end{equation}
and then perform a conformal transformation to another set of null
coordinates, $x^\pm=\tan{\xi^\pm}$, to bring infinity to a finite
coordinate distance.  The final step in constructing the Penrose
diagram in figure~\ref{pen_normal}b is to identify which values of 
$\xi^+$ and $\xi^-$ correspond to physical values 
$-\infty  < \eta\leq 0$ and $0 \le r < \infty$.  
The initial singularity is located at past timelike 
infinity and at past null infinity.

The Penrose diagrams in figure~\ref{pen_normal} are constructed from
classical solutions of the gravitational equations.  The difference
between the two diagrams can be traced to the behavior of the scale
factor at very early times.  In fact, the divergent contribution 
to the past conformal time in (\ref{diveta}) comes from within a
Planck time following the initial singularity.
Quantum effects will presumably dominate during this period and
classical solutions are unlikely to describe the physics correctly.
The physical relevance of the Penrose diagram in 
figure~\ref{pen_normal}b is therefore questionable and in 
section~\ref{entropy} we will indeed find that such diagrams 
represent behavior that is inconsistent with holography and the
covariant entropy bound.  The appropriate way to deal with models
with $\beta\geq 1$ and $\kappa\leq 0$ is to put a cutoff at the
Planck time on the lower bounds of integration in (\ref{diveta})
to avoid extending the classical description into a period 
dominated by quantum effects \cite{kallin}.  The Penrose diagram
then becomes that of figure~\ref{pen_normal}a.

\subsection{Radiation in 3+1 dimensions}

We now consider the $3{+}1$-dimensional big-bang solutions with radiation 
(\ref{radscale}), for which 
\begin{equation}
\label{conftime}
\tilde{\eta} = \vert\kappa\vert^{1/2}
\int_0^{\tau} {dx\over 
\sqrt{\sinh{2x}+(\kappa/2)(1-\cosh{2x})}} \, .
\end{equation}
The models with $\kappa<2$ accelerate forever in comoving time but the
conformal time nevertheless remains finite in the limit 
$\tau\rightarrow\infty$.  The maximal conformal time is given by
\begin{equation}
\tilde{\eta}_{max} = {2\vert\kappa\vert^{1/2}\over\sqrt{2-\kappa}}\,
K\left[-{2+\kappa\over 2-\kappa}\right] ,
\label{etamax}
\end{equation}
where $K[m]=\int_0^1 dy\,\{(1-y^2)(1-my^2)\}^{-1/2}$ is the
complete elliptic integral of the first kind\footnote{The fact that $K[-1] 
= \frac{1}{4} B\left(\frac{1}{2},\frac{1}{4}\right)$
provides a check on our results. Otherwise (\ref{eulerb}) and (\ref{etamax})
would give conflicting values for $\eta_{max}$ in the $3{+}1$-dimensional
radiation model with $\kappa =0$.}.
Conformal time also remains finite in a re-collapsing universe with 
$\kappa>2$.  In this case the integration in (\ref{conftime}) is cut off at 
the big crunch at $\tau_f={1\over 2}\log ({\kappa+2\over\kappa-2})$.

The global causal structure depends on the shape of the Penrose diagram,
which in turn depends on the value of the maximal conformal time.  
For solutions with $\kappa>0$ the Penrose diagram is 
`tall' if $\tilde{\eta}_{max}>\pi$.  In this case the entire 
spatial geometry is eventually in the causal past of any given comoving 
observer.  If, on the other hand, $\tilde{\eta}_{max}<\pi$ the geometry 
is said to be `short' and a comoving observer can only be influenced by 
events in part of the spatial geometry.

\FIGURE{
    \psfrag{a}{(a)}
    \psfrag{b}{(b)}
    \psfrag{c}{(c)}
    \psfrag{d}{(d)}
    \epsfig{file=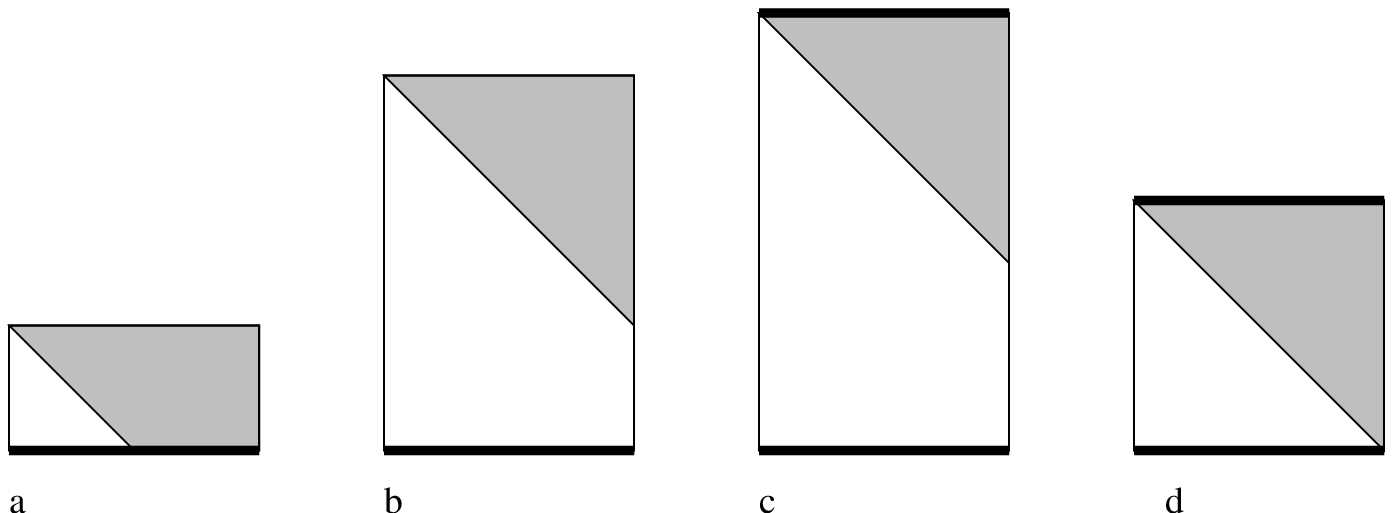, width=13cm}
    \caption{Penrose diagrams for $\kappa>0$ models.
                Thick lines indicate singularities.
                (a)~Short big-bang geometry, (b)~tall big-bang geometry, 
        (c)~tall re-collapsing geometry,
        (d)~marginally tall re-collapsing geometry.}
    \label{pen_multi}
}

The family of big-bang solutions with $0<\kappa<2$ is easily seen to 
include both tall and short geometries, as indicated in 
figure~\ref{pen_multi}.  
In the limit $\kappa\rightarrow 2$ the solution approaches the Einstein 
static universe and $\tilde{\eta}_{max}$ in (\ref{etamax}) diverges 
logarithmically, leading to an arbitrarily tall Penrose diagram.  
On the other hand, $\tilde{\eta}_{max}$ goes to zero when 
$\kappa\rightarrow 0$ and the associated Penrose diagram becomes 
arbitrarily short.  The $\kappa=0$ Penrose diagram in 
figure~\ref{pen_normal}a may be obtained from the $\kappa>0$ diagrams by 
rescaling both axes in figure~\ref{pen_multi}a by $1/\sqrt{k}$ before 
taking the $k\rightarrow 0$ limit.  
The Penrose diagrams for hyperbolic $\kappa<0$ big-bang geometries are 
the same as for the spatially flat case and are shown in 
figure~\ref{pen_normal}. 

For the re-collapsing big-bang solutions with $\kappa>2$, conformal time 
is cut off at the big crunch singularity.  
These geometries are nevertheless all tall.  For $\kappa$ close to 2 they 
are very tall (figure~\ref{pen_multi}c) as the lifetime of the universe 
diverges in the $\kappa\rightarrow 2^+$ limit, while for large $\kappa$ 
(figure~\ref{pen_multi}d) they are marginally tall 
($\tilde{\eta}_{max}\rightarrow\pi$) and we recover the 
behavior of closed cosmological models with $\Lambda=0$ where the spatial 
geometry becomes fully visible to a comoving observer at the 
big crunch.

The causal structure of the $\kappa>2$ bounce solutions (\ref{bounce})
can be analyzed in a similar fashion.  In this case we have 
$-\tilde{\eta}_{max}\leq\tilde{\eta}\leq\tilde{\eta}_{max}$ with
\begin{eqnarray}
\tilde{\eta}_{max} &=& \sqrt{\kappa}\int_0^\infty 
{dx\over \sqrt{{\kappa\over 2}+\sqrt{{\kappa^2\over 4}-1}\cosh 2x}} 
\nonumber \\
&=&2\int_0^1 {dy\over\sqrt{2y^2+\sqrt{1-{4\over\kappa^2}}(1+y^4)}}\,.
\end{eqnarray}
The integral is clearly a decreasing function of $\kappa$, with 
$\tilde{\eta}_{max}\rightarrow {\pi\over 2}$ as 
$\kappa\rightarrow\infty$.  This means that the bounce geometries 
are all `tall', in agreement with a general result of Gao and 
Wald \cite{gaowald} regarding asymptotically de~Sitter 
spacetimes.\footnote{In contrast, some of our big-bang geometries
were `short'.  This does not contradict the results of Gao and Wald
since these are singular spacetimes which are only future 
asymptotically de~Sitter.}  The bounces with $\kappa\rightarrow\infty$
are approaching de~Sitter spacetime and are therefore only 
marginally tall but $\tilde{\eta}_{max}$ diverges as 
$\kappa\rightarrow 2^+$ and so the family of bounce solutions 
contains geometries with arbitrarily tall Penrose diagrams, 
see figure~\ref{pen_bounce}.
\FIGURE{
    \psfrag{a}{(a)}
    \psfrag{b}{(b)}
    \epsfig{file=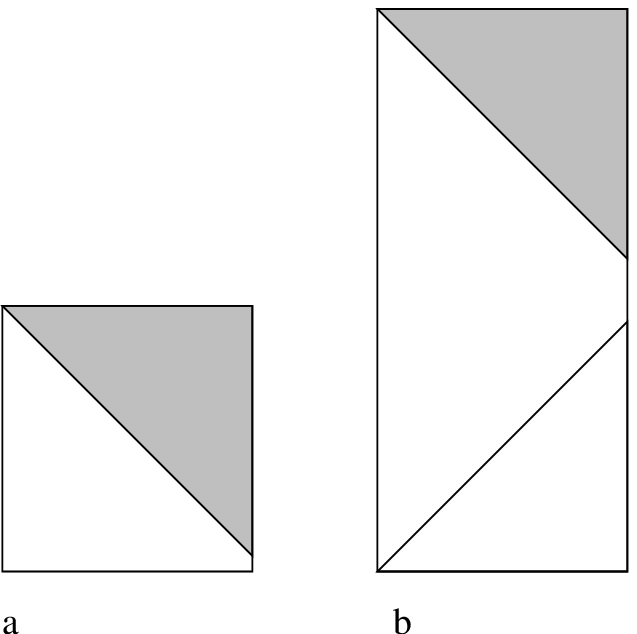, width=5.5cm}
    \caption{(a)~Marginally tall bounce geometry, 
        (b)~very tall bounce geometry.\\ }
    \label{pen_bounce}
}

For $\kappa>0$ solutions the line element (\ref{dless}) determines the 
proper volume of the spatial geometry as a function of comoving time,
\begin{equation}
V(\tau) = {2\pi^2\over \kappa^{3/2}}\,u(\tau)^3 
\left({3\over\Lambda}\right)^{3/2} \,,
\end{equation}
in 3+1 spacetime dimensions (with a corresponding formula for the 
volume of $4{+}1$-dimensional dust universes).
The minimum volume of a bounce universe occurs at $\tau=0$,
\begin{equation}
V_{min}= \left({1\over 2}+\sqrt{{1\over 4}-{1\over\kappa^2}}\right)^{3/2}
2\pi^2 \left({3\over\Lambda}\right)^{3/2} \,.
\label{minvolume}
\end{equation}
It is an increasing function of $\kappa$, which approaches the minimum 
spatial volume of de~Sitter space in global coordinates for large $\kappa$, 
and varies over a relatively narrow range:  
$V_{min}(\kappa\rightarrow \infty) = V_{min}(dS)= 
2^{3/2}\,V_{min}(\kappa\rightarrow 2)$.

\subsection{Dust models}
We now turn our attention to models with 
matter in the form of pressureless dust.  As mentioned previously, the 
$4{+}1$-dimensional case is identical the $3{+}1$-dimensional radiation models
we have just discussed.  For $3{+}1$-dimensional models with dust we do not 
have explicit analytic solutions at our disposal, except for the 
$\kappa=0$ model, but numerical results are in qualitative agreement 
with the picture obtained in 4+1 dimensions.  In particular, 
3+1-dimensional bounce geometries with dust are all tall, as are
the re-collapsing big-bang geometries.  The same numerical calculations
apply to 2+1-dimensional models with radiation.

In section~(\ref{twoddust}) we found explicit solutions for dust models
in 2+1 dimensions.   The conformal time of the bounce geometries 
(\ref{twodbounce}) is given by
\begin{eqnarray}
\tilde{\eta}(\tau) &=& \sqrt{\kappa\over \kappa -1} 
\int_0^\tau {dx\over \cosh{x}}  \nonumber \\
&=& \sqrt{\kappa\over \kappa -1}
\left(2\arctan{(e^\tau)} - {\pi\over 2}\right) \,.
\end{eqnarray}
The range is $-\tilde{\eta}_{max}<\tilde{\eta}<\tilde{\eta}_{max}$ with
\begin{equation}
\tilde{\eta}_{max}=\sqrt{\kappa\over \kappa -1}\, {\pi\over 2} \,.
\end{equation}
As $\kappa$ is varied over its allowed range the bounce geometries go
from being very tall as $\kappa\rightarrow 1^+$ to marginally tall
in the $\kappa\rightarrow\infty$ pure de~Sitter limit.
\FIGURE{
    \psfrag{eta}{\footnotesize{$\tilde{\eta}=0$}}
    \psfrag{chi0}{\footnotesize{$\chi=0$}}
    \psfrag{chipi}{\footnotesize{$\chi=\pi$}}
    \epsfig{file=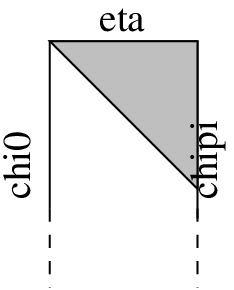, width=4cm}
    \caption{Big-bang geometry for dust in 2+1 dimensions, with 
        $0<\kappa<1$.}
    \label{pen_past}
}

So far the story is similar to the higher dimensional models but when it 
comes to the big-bang models (\ref{twodbb}) we find different behavior.
Conformal time is logarithmically divergent as $\tau\rightarrow 0$ for 
these models,
\begin{eqnarray}
\tilde{\eta}(\tau) &=& -\sqrt{\vert\kappa\vert\over 1-\kappa}
\int_\tau^\infty {dx\over \sinh{x}}  \nonumber \\
&=& \sqrt{\vert\kappa\vert\over 1-\kappa}
\log{\left({e^\tau -1\over e^\tau +1}\right)} \,.
\end{eqnarray}
The Penrose diagram for big-bang models with $0<\kappa<1$ is shown in
figure~\ref{pen_past}.  The dimensionless radial variable has finite 
range $0\leq\chi\leq\pi$, while the dimensionless conformal time 
ranges from $\tilde{\eta}\rightarrow -\infty$ at the initial 
singularity to $\tilde{\eta}=0$ in the asymptotic future.  It follows 
that the entire spatial geometry is in the causal past of an observer
at $\chi=0$ at any finite comoving time $\tau>0$.  We note, however,
that this strange property is derived from a classical solution but
comes from a period of early evolution following an inital singularity
where classical solutions have limited validity.

The big-bang geometry with $\kappa= 0$ is an example of a spatially 
flat model with $\beta\geq 1$ (in fact $\beta=1$) and the Penrose 
diagram is shown in figure~\ref{pen_normal}b.  The same Penrose 
diagram applies to the hyperbolic $\kappa<0$ models and the same 
reservations apply concerning the physical relevance of such diagrams.

\subsection{Negative pressure models}
\label{nppenr}

We close this section by analyzing the causal structure of the
cosmological models with negative pressure matter introduced in
section~(\ref{negpress}).  Let us first consider the big-bang
models (\ref{npbb}) with $\kappa<0$.  The initial expansion is
linear in $\tau$, just as it was for dust models in 2+1 dimensions
and the construction of the Penrose diagram is analogous.
Conformal time is defined 
\begin{eqnarray}
\tilde{\eta}(\tau) &=& -\sqrt{-\kappa}
\int_\tau^\infty {dx\over \sinh^2{x\over 2}+\sqrt{-\kappa}\sinh{x}}  
\nonumber \\
&=& \log{\left({e^\tau -1\over e^\tau -b}\right)} \,,
\end{eqnarray}
with $b={1-2\sqrt{-\kappa}\over 1+2\sqrt{-\kappa}}$.  The 
conformal time is logarithmically divergent as $\tau\rightarrow 0$
and the Penrose diagram of the classical geometry is identical to the 
one in figure~\ref{pen_normal}b.

For $\kappa>0$ we have the bounce solutions (\ref{npbounce}) 
with conformal time given by
\begin{eqnarray}
\tilde{\eta}(\tau) &=& \sqrt{\kappa}
\int_0^\tau {dx\over \sqrt{\kappa+{1\over 4}}\cosh{x}-{1\over 2}}
\nonumber \\
&=& 2\left(\arctan{g(\tau)}-\arctan{g(0)}\right) \,,
\end{eqnarray}
where $g(\tau)=\sqrt{1+{1\over 4\kappa}}\,e^\tau - \sqrt{1\over 4\kappa}$.  
The range is $-\tilde{\eta}_{max}<\tilde{\eta}<\tilde{\eta}_{max}$ with
\begin{equation}
\tilde{\eta}_{max}=\pi-2\arctan{g(0)}.
\end{equation}
Since $\tilde{\eta}_{max}>\pi/2$ for all $\kappa>0$ these bounce solutions
are all tall.  The limiting behavior is $\tilde{\eta}_{max}\rightarrow\pi/2$
as $\kappa\rightarrow\infty$, as appropriate for a bounce solution that
approaches the de~Sitter vacuum, and $\tilde{\eta}_{max}\rightarrow\pi$
as $\kappa\rightarrow 0$.  The second limit is interesting in that a 
bounce geometry with $\tilde{\eta}_{max}\geq\pi$ is not only tall but 
`very tall' in the terminology of \cite{lmm}.  Then an entire Cauchy 
surface lies in the causal past of a comoving observer at late times
and also in the causal future of the same observer at an early 
enough time.  For this Cauchy surface we can, for example, choose the 
spatial slice at $\tau=0$, which is also when the proper spatial volume
of the bounce universe takes its minimum value,
\begin{equation}
V_{min}=
\left(\sqrt{1+{1\over 4\kappa}} - \sqrt{1\over 4\kappa}\right)^{n} 
V_{min}(dS) \,,
\end{equation} 
where $V_{min}(dS)=a_n[n(n-1)/ 2\Lambda]^{n/2}$ is the minimum
volume of the spatial slice of $n{+}1$-dimensional de~Sitter spacetime
in global coordinates.  We note that the minimum volume of the bounce 
universe vanishes as $\kappa^{n}$ in the $\kappa\rightarrow 0$ limit.

We also have a $\kappa =0$ big-bang solution, which is a special case
of (\ref{zerokappa}) with $\beta =2$, and has infinite spatial volume
at any $\tau >0$. The conformal time is linearly divergent as $\tau
\rightarrow 0$, rather than the logarithmic divergence found for
$\kappa <0$, but the Penrose diagram of the classical solution remains 
that of figure~\ref{pen_normal}b.

\section{Cosmological horizons}
\label{horizons}

A finite maximal conformal time implies a cosmological event horizon 
for a comoving observer.  This is, for example, evident in the Penrose 
diagram in figure~\ref{pen_normal}a where there are regions of spacetime 
which can have no influence on an observer at $\chi=0$.  
The existence of a horizon has interesting implications for cosmological
observations in an accelerating universe \cite{gbehg} and also
motivates questions concerning holography and entropy bounds, which
we address in section~\ref{entropy}.
The event horizon forms the boundary of the causal past of a comoving 
observer in the asymptotic future.  It can therefore only be described 
from knowledge of the full future evolution of the cosmological model. 
The shaded regions in the various Penrose diagrams displayed in 
section~\ref{confdiagrams} are outside the event horizon of an observer
at $\chi=0$.

There is another notion of horizon which is based on local data.  This
is the apparent horizon, which is at the boundary of future trapped
(or anti-trapped) spatial regions at any given cosmic time.  In the
de~Sitter vacuum the apparent horizon coincides with the event horizon,
but in an evolving geometry, which is future asymptotically 
de~Sitter, the two will only agree in the far future.  

\subsection{The event horizon}
Consider $n{+}1$~dimensional cosmological models that are future
asymptotically de~Sitter.  The scale factor of these solutions
grows exponentially at late times.  In terms of our dimensionless 
variables the asymptotic rate of expansion is always the same,
\begin{equation}
u(\tau)\approx a\, e^\tau \,,
\label{uasymptotic}
\end{equation}
but subleading behavior, such as the constant $a$, in general
depends on the number of dimensions, the equation of state, and the 
matter energy density.

The line element (\ref{dless}) implies that the event horizon of an 
observer at $\chi=0$ is the null surface 
\begin{equation}
\chi_{eh}(\tau) = \tilde{\eta}_{max} -\tilde{\eta}(\tau) \,.
\label{horizon}
\end{equation}
For a tall big-bang geometry with $\kappa >0$, for which 
$\tilde{\eta}_{max}-\tilde{\eta}_{min}>\pi$, the
event horizon comes into existence at $\tau=\tau_0$, at which 
$\chi_{eh}(\tau_0)=\pi$, while for a short $\kappa >0$ geometry, 
and also for all big-bang
models with $\kappa\leq 0$, the event horizon forms at the initial
singularity.  

The proper area of the intersection of the event horizon with the spatial
volume at comoving time $\tau$ is given by
\begin{equation}
{\cal A}_{eh}(\tau) = a_{n-1}
\left(\vert\kappa\vert^{-1/2}u(\tau) f(\chi_{eh}(\tau))\right)^{n-1}
\left({n(n-1)\over 2\Lambda}\right)^{(n-1)/2} \,,
\label{horizonarea}
\end{equation}
where $f(\chi)$ depends on the sign of $\kappa$ as in (\ref{ffunc}).
The area starts out at zero when the horizon forms and 
then increases monotonically with time in all our models that are
future asymptotically de~Sitter.
It is straightforward to show, using the asymptotic form 
(\ref{uasymptotic}), that 
\begin{equation}
\vert\kappa\vert^{-1/2}u(\tau) f(\chi_{eh}(\tau))
\rightarrow 1 \qquad {\rm as} \qquad \tau\rightarrow\infty \,,
\end{equation}
for all $a$.  The limiting value of the horizon area is therefore 
independent of $\kappa$,
\begin{equation}
\label{dsarea}
\lim_{\tau\rightarrow\infty} {\cal A}_{eh}(\tau) = a_{n-1}
\left({n(n-1)\over 2\Lambda}\right)^{(n-1)/2}
= {\cal A}_{dS}\,.
\end{equation}
As expected, the area approaches the area of the cosmological horizon 
in empty $n{+}1$-dimensional de~Sitter spacetime with cosmological constant 
$\Lambda$.  As time goes on in an accelerating universe the spatial region 
inside the event horizon occupies an ever smaller patch around $\chi=0$ 
in comoving coordinates and the global structure of the spatial geometry 
becomes irrelevant.  

It is also easy to see, after appropriate rescaling of variables by powers 
of $k$, that the $\kappa\rightarrow 0$ limit of ${\cal A}_{eh}(\tau)$ is 
smooth and gives the same result as a direct $k=0$ calculation.  
\FIGURE{
    \psfrag{A}{${\cal A}_{eh}/{\cal A}_{dS}$}
    \psfrag{tau}{$\tau$}
    \epsfig{file=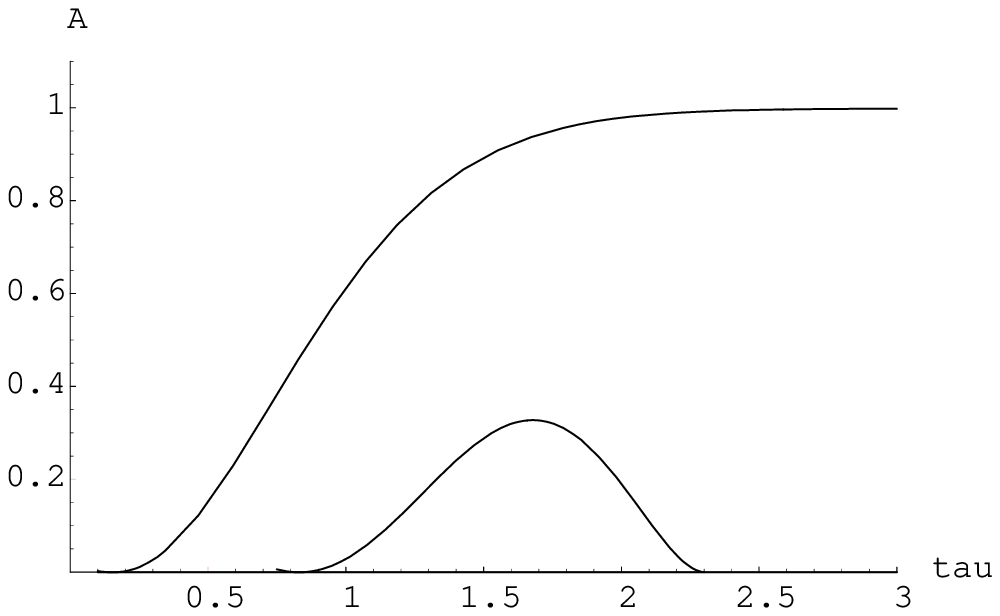, width=5.5cm}
    \caption{The area of the event horizon relative
            to the corresponding de~Sitter horizon area, as a
        function of $\tau$ for ever-expanding and re-collapsing
        big-bang geometries.}
    \label{Ah}
}

Now consider the bounce solutions which are both past and future 
asymptotically de~Sitter.   The definition of the event horizon in equation 
(\ref{horizon}) continues to hold, and since these geometries are all 
tall the event horizon comes into existence at $\chi_{eh}(\tau_0)=\pi$ at 
some finite comoving time $\tau_0$.  
The area starts out at zero, then increases with time until it 
approaches ${\cal A}_{dS}$ at late times.  Note that the area of the 
event horizon grows with time even when the horizon forms at $\tau_0<0$ 
in the contracting phase of the bounce evolution. 

Finally, we come to the big-bang solutions which re-collapse in
a big-crunch singularity.  In this case, we can define the event
horizon of an observer at $\chi=0$ as in equation (\ref{horizon}) with 
$\tilde{\eta}_{max}$ taken as the conformal time at the final singularity.
With this definition the event horizon starts with zero area at 
$\chi=\pi$ at some time after the big bang.  The horizon area grows 
initially but as the geometry collapses towards the big crunch the 
area shrinks again and goes to zero at the final singularity.
Figure~\ref{Ah} shows how the ratio ${\cal A}_{eh}(\tau)/{\cal A}_{dS}$ 
evolves for two big-bang models, one that accelerates forever and another
that re-collapses.

\subsection{The apparent horizon}

Consider spherical $n{-}1$~surfaces centered on an observer at $\chi=0$ in 
one of our $n{+}1$-dimensional cosmological models.  Such a surface is
uniquely characterized by a value of $\chi$ and a conformal time 
$\tilde{\eta}$.  A radial null ray is orthogonal to such a spherical 
surface.  It is future (past) directed if $\tilde{\eta}$ 
increases (decreases) along the ray away from the sphere and outgoing
(incoming) if $\chi$ increases (decreases).  Thus there are four 
families of null rays orthogonal to each sphere. 

Let $\lambda$ be the affine parameter of null rays in one of the 
null directions orthogonal to a given spherical surface.  By a linear
transformation of $\lambda$ on each null ray we can set $\lambda=0$
where this family of null rays intersects our surface and also ensure that
we advance at the same rate along all the null rays at $\lambda=0$.
The surface intersected by our null rays at infinitesimal parameter 
distance $d\lambda$ is then also spherical.  The expansion $\theta$
is then defined in terms of the rate of change of the proper area 
(\ref{properarea}) of the surface intersected by the family of null
rays,
\begin{equation}
\label{expansiondef}
\theta(\lambda)={1\over {\cal A}}\,{d{\cal A}\over d\lambda}\,.
\end{equation}
By a more general construction the expansion can be defined locally
on any $n{-}1$ surface~\cite{hawell}, but for the purposes of the 
present paper the above definition involving spherical surfaces will be 
sufficient.  

If a family of null rays orthogonal to a given spherical surface has
non-positive expansion, $\theta(\lambda)\leq 0$, it is referred to as
a light-sheet of that surface.  At least two of the four families of 
null rays orthogonal to a surface will satisfy this condition.  
The light-sheet extends along the family of null rays until positive
expansion is encountered, in which case it terminates.  Light sheets
play a key role in the covariant entropy bound \cite{boussoreview} which we 
will apply to some of our cosmological models in section~\ref{entropy}.

In a normal region of spacetime outgoing (incoming) future directed
null rays orthogonal to a surface have positive (negative) expansion.  
If both future directed families of null rays have negative expansion, 
{\it i.e.} are light-sheets, the surface is said to be in a future trapped 
region.  This occurs, for example, inside the event horizon of a 
Schwarzschild black hole.  If, on the other hand, both future directed 
families of null rays have positive expansion, {\it i.e.} the past 
directed families are light-sheets, the spherical surface is in a future 
anti-trapped region.  An example of this behavior is provided by 
spherical surfaces outside the cosmological event horizon of an observer 
in de~Sitter spacetime.

The boundary between normal and trapped (or anti-trapped) regions is 
called an apparent horizon.  At least one pair out of the four families of
null rays has vanishing expansion there.  In de~Sitter spacetime, 
for example, there is an apparent horizon which coincides with the 
de~Sitter event horizon.  There is no reason to expect the two types
of horizons to coincide in more general, evolving geometries.

In our cosmological solutions we can, for example, identify the affine 
parameter with the conformal time itself, 
$\lambda=\pm(\tilde{\eta}-\tilde{\eta}_0)$, where
$\tilde{\eta}_0$ is the conformal time coordinate of our spherical
surface and the sign depends on whether the family of null rays is future
or past directed.  Null rays that are orthogonal to the surface are
radial and satisfy $d\chi/d\lambda=\pm 1$, with the sign indicating
whether they are outgoing or incoming.  For future directed
null rays the expansion is thus given by
\begin{eqnarray}
\theta &=& (n-1)\left({1\over u}{du\over d\tilde{\eta}}
\pm {1\over f}{df\over d\chi}\right) \nonumber \\
&=&  (n-1)\left({1\over\sqrt{\vert\kappa\vert}}{du\over d\tau}
\pm {1\over f}{df\over d\chi}\right) \,,
\end{eqnarray}
with $f(\chi)$ given in (\ref{ffunc}).
An apparent horizon is then located wherever
\begin{equation}
{1\over \vert\kappa\vert}\left({du\over d\tau}\right)^2
= {1\over f^2}\left({df\over d\chi}\right)^2 \,.
\label{horloc}
\end{equation}
In a universe undergoing accelerated expansion the apparent horizon 
separates a normal region at $0\leq\chi<\chi_{ah}$ and a future 
anti-trapped region at $\chi>\chi_{ah}$.
When $\kappa>0$ we have $f(\chi)=\sin{\chi}$ and the apparent
horizon condition (\ref{horloc}) can be rewritten as 
\begin{equation}
\label{ahorizon}
\sin{(\chi_{ah}(\tau))}
=\left(1+{1\over \kappa}\left(du/d\tau\right)^2\right)^{-1/2} \,.
\end{equation}
The area of the apparent horizon is then given by
\begin{eqnarray}
{\cal A}_{ah}(\tau) &=& a_{n{-}1}
\left(\kappa^{-1/2}u(\tau)\sin{(\chi_{ah}(\tau))}\right)^{n-1}
\left({n(n{-}1)\over 2\Lambda}\right)^{(n-1)/2} \nonumber  \\
&=&  \left({u^2\over\kappa+(du/d\tau)^2}\right)^{(n-1)/2}
{\cal A}_{dS}  \,,
\label{eitthvad}
\end{eqnarray}
where ${\cal A}_{dS}$ is the area of the corresponding de~Sitter event 
horizon.  By using the evolution equation (\ref{dusquared}) we can 
eliminate the derivative of the scale factor to obtain
\begin{equation}
\label{aharea}
{\cal A}_{ah}(\tau)
=\left(1+u^{-(1+\alpha)n}\right)^{-(n-1)/2}{\cal A}_{dS} \,.
\end{equation}
In the corresponding calculations for $\kappa<0$ solutions, equation
(\ref{ahorizon}) is replaced by
\begin{equation}
\sinh{(\chi_{ah}(\tau))}
=\left(-1+{1\over \vert\kappa\vert}
\left(du/d\tau\right)^2\right)^{-1/2} \,,
\end{equation}
and for the spatially flat case one has $r_{ah}(t) =(dR/dt)^{-1}$.  
In all cases the end result in (\ref{eitthvad}) and (\ref{aharea}) for
the area of the apparent horizon is the same.  
We note that ${\cal A}_{ah}<{\cal A}_{dS}$ for all physical values of 
the scale factor and that ${\cal A}_{ah}$ tends towards 
${\cal A}_{dS}$ as $u\rightarrow\infty$.  

An interesting feature of $\kappa>0$ solutions, whose spatial sections
are $n$-spheres, is that the apparent horizon condition (\ref{ahorizon})
has two solutions that are equidistant from the equator.  In an 
expanding $\kappa>0$ universe the apparent horizon in the northern
hemisphere is the boundary between the normal region near the observer
at $\chi=0$ and an anti-trapped region where both future directed
families of null rays have positive expansion.  Beyond the other apparent 
horizon, which is in the southern hemisphere, we are again in a normal 
region where one future directed family of null rays has negative
expansion and the other positive.  It is, however, the null rays that
are outgoing with 
respect to the observer at $\chi=0$ that have negative expansion.
\FIGURE{
    \psfrag{t}{$\tau=0$}
    \epsfig{file=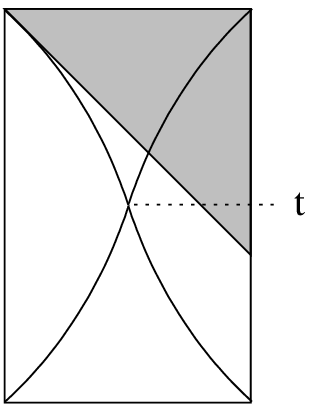, width=4cm}
    \caption{Penrose diagram for a bounce solution with the event 
    horizon and both apparent horzions indicated.}
    \label{pen_ah}
}
Now consider bounce solutions. At the minimal scale factor we have
$du/d\tau=0$.  It then follows from equation (\ref{ahorizon}) that both 
apparent horizons must be at the equator at that point.  The location
of the two apparent horizons in comoving coordinates can be traced 
throughout the bounce evolution as follows.  In the asymptotic past a
pair of apparent horizons emerges from the north and south poles and
in the contracting $\tau<0$ phase the two polar regions are normal
while the region between the apparent horizons is future trapped.  
As the scale factor approaches its minimum value the trapped region 
contracts to a narrow band around the equator, which then vanishes at 
$\tau=0$ when the two apparent horizons meet.  In the expanding 
$\tau>0$ phase the apparent horizons separate again but now the 
equatorial region has become anti-trapped.  In the asymptotic future
the two apparent horizons approach the north and south poles.

The cosmological event horizon in a future asymptotically de~Sitter
spacetime is an example of a family of null rays that are orthogonal to 
transverse $n{-}1$ spheres.  The area of the transverse sphere 
intersected by the event horizon increases with comoving time. It 
follows that the past directed null rays along the event horizon form a 
light-sheet and also that $\chi_{eh}(\tau)\geq \chi_{ah}(\tau)$, 
{\it i.e.} the event horizon is always outside the apparent horizon.
In a $\kappa>0$ geometry this statement refers to the apparent horizon 
that is closer to $\chi=0$.  The event horizon may or may not be outside 
the other apparent horizon at $\chi = \pi-\chi_{ah}(\tau)$.  Note, 
furthermore, that an event horizon that forms at $\chi>\pi/2$ in a 
$\kappa>0$ geometry may have less area than the apparent horizon early 
on, despite being outside the apparent horizon.  To give an example, 
figure~\ref{pen_ah} shows the location of the event horizon and both
apparent horizons in the Penrose diagram for a bounce geometry.

\section{Cosmological evolution as renormalization group flow}
\label{cfunct}

The $n{+}1$-dimensional de~Sitter vacuum is conjectured to be dual
to an $n$-dimensional conformal field theory \cite{strominger}.
The case is strongest for three-dimensional spacetime where the
dual field theory is two-dimensional.  An analysis of the asymptotic
symmetry generators of three-dimensional de~Sitter spacetime
\cite{strominger} revealed a conformal algebra with central charge
\begin{equation}
c={3\over 2G\Lambda} \,.
\end{equation}
On dimensional grounds, the expected generalization to higher
dimensions is
\begin{equation}
\label{ccharge}
c\sim {1\over G\Lambda^{(n-1)/2}} \,,
\end{equation}
up to an undetermined constant of proportionality.

More generally in this context, cosmological evolution in a model which 
is future asymptotic to de~Sitter spacetime represents a reverse 
renormalization group flow on the dual field theory
side \cite{strominger2,vijay}.  A candidate c-function was proposed,
that could be evaluated on the gravity side and shown to decrease
towards the infrared along the renormalization group flow, provided
the matter in the cosmological model satisfies the weak energy
condition, {\it i.e.} $\rho\geq -P$ and $\rho>0$.  The original
papers \cite{strominger,vijay} considered spatially flat cosmological
models with $k=0$ but their c-function was subsequently
generalized \cite{lmm} so as to apply to $k\neq 0$ models as well.

We take a different route to arrive at the same c-function as
\cite{lmm}.  Our starting point is the observation that the central
charge (\ref{ccharge}) of the $n$-dimensional fixed point theory is
proportional to the area of the event horizon in Planck units in
the corresponding $n{+}1$-dimensional de~Sitter spacetime.  We wish
to identify a c-function that reduces to this central charge at the
ultraviolet fixed point of the renormalization group flow and
decreases along the flow towards the infrared, {\it i.e.} increases
as the cosmological scale factor grows and approaches de~Sitter
expansion.  The evolving horizon area is then a natural candidate
for a c-function, but we have a choice to make between the area of
the event horizon and that of the apparent horizon.  At least two
factors count against choosing the the area of the event horizon.
One is that global information about the spacetime geometry is
required in order to determine the location of the event horizon,
and therefore also its area, at a given comoving time.  The other
is the related fact that for some geometries there are times when
the event horizon has yet to come into existence.  The apparent
horizon, on the other hand, is defined by local data on any spatial
slice and it exists at all times in all our cosmological models.

We therefore define the c-function associated with an asymptotically
de~Sitter spacetime to be (proportional to) the area of the
cosmological apparent horizon in Planck units,
\begin{equation}
\label{cfunction}
c\sim {{\cal A}_{ah}\over G} \,.
\end{equation}
With this definition we have a prescription for evaluating the
c-function on any constant time slice, and the c-theorem becomes a
geometric statement about the increase in area of the apparent
cosmological horizon in an accelerating universe.

For any metric of Robertson-Walker form (\ref{rwmetric}),
our definition gives 
\begin{equation}
c \sim \left({u^2\over\kappa+(du/d\tau)^2}\right)^{(n-1)/2}
       \frac{{\cal A}_{dS}}{G}
\end{equation}
which is the same c-function as the one advocated in \cite{lmm}.  
Our proposal can be viewed as providing a geometric interpretation of 
their c-function.  The definition of an apparent horizon can be extended 
to anisotropic cosmological models and we expect our geometric 
approach to remain useful for such models as well.

By using the evolution equation (\ref{dusquared}) the c-function for 
our models may be written,
\begin{equation}
c \sim \frac{{\cal A}_{dS}}{\left(1+u^{-(1+\alpha)n}\right)^{(n-1)/2}G}.
\end{equation}
This expression makes manifest the UV/IR correspondence of the proposed 
dS/cft duality.  The c-function only depends on the overall scale $u$ of 
the geometry and $c$ decreases as we move to smaller scales on the gravity 
side, which corresponds to flowing to the infrared in the dual field 
theory.  

This means, for example, in a big-bang model approaching de~Sitter 
acceleration in the future that it is the reverse of the cosmological
evolution that corresponds to renormalization group trajectories starting at 
an ultraviolet fixed point theory.  The c-function then decreases 
monotonically along the flow towards the infrared but it is less clear how 
to interpret the endpoint of the flow where the c-function goes to zero 
at a big-bang singularity on the gravitational side. 
Evolution forward in time corresponds to renormalization group flow in
a big-crunch model that is past asymptotically de~Sitter, with the 
c-function again vanishing at the singular endpoint.  
Re-collapsing big-bang geometries are singular at both ends and do not
lend themselves to straightforward interpretation as renormalization 
group flow.  For bounce solutions, the gravitational evolution is 
non-singular everywhere but instead the scale factor is not a monotonic 
function of comoving time.  If we follow 
the cosmological evolution backwards (or forwards) in time the scale factor 
eventually stops decreasing and begins to grow, and it is clear that the
full time history cannot represent a single renormalization group flow in 
the usual sense.  In \cite{lmm}, where analogous bounce geometries 
were considered, it is suggested to view them as matching two separate 
renormalization group flows arriving at the same effective field theory
from opposite time directions.  
 
\section{Holography and entropy}
\label{entropy}

We end the paper with an application of the holographic principle 
\cite{thooft,lenny}, in the form of the covariant entropy 
bound \cite{boussoreview}, in the context of these cosmological models.
The basic idea is to use the fact that the cosmological event horizon of 
a comoving observer in a future asymptotically de~Sitter spacetime
provides 
a past directed light-sheet for any transverse $n{-}1$-sphere it
intersects.  
We will apply the covariant entropy bound to the light-sheet of such a 
transverse sphere at asymptotically late time.  In this case the area
approaches the area of the corresponding de~Sitter horizon ${\cal A}_{dS}$, 
and we can compare the total entropy that crosses the light-sheet to 
${\cal A}_{dS}/4$.
We will make the assumption that entropy in a cosmological spacetime
may be described by a local `entropy fluid'.  This cannot be correct in
any fundamental sense, given that entropy is not a local quantity, but
it is an approximation often made in cosmology.

Recall that the area of the de~Sitter horizon in (\ref{dsarea}) involves 
an inverse power of $\Lambda$ and therefore becomes large in the 
limit of small $\Lambda$.  On the other hand, for models with $\kappa\geq 0$
the volume enclosed by this area grows even faster with vanishing 
$\Lambda$ and one might worry that this could lead to a violation of the 
entropy bound  for sufficiently small values of the cosmological constant.  
As we will see below, such a conflict only arises in models with somewhat 
exotic matter content and there it can be traced to the failure of 
classical solutions to correctly describe the physics close to the
initial singularity.

Let us first apply these ideas to cosmological models with matter in 
the form of radiation, under the assumption that the cosmological 
expansion is adiabatic.  In this case, no violation of the entropy bound
is found but the exercise serves to illustrate the argument.  A simple 
relationship can be established between the proper entropy density $s$ 
and the energy density $\rho$ at a given comoving time.  Both these 
quantities are related to the temperature $T$ of the gas of radiation,
\begin{equation}
\rho \sim T^{n+1} \,,\qquad s\sim T^n \,.
\end{equation}
The energy density can in turn be related to the cosmological constant 
and the scale factor $u(\tau)$ through equation (\ref{rhorad}), leading to
\begin{equation}
s \sim \rho^{n/(n+1)}\sim u^{-n}\left({\Lambda\over G}\right)^{n/(n+1)}\,.
\end{equation}
The dependence on the scale factor reflects the fact that the entropy 
density is diluted by the cosmological expansion, while the comoving
entropy density $\tilde{s}\equiv u^n\, s$ remains constant throughout 
the evolution.  

The next step is to obtain the total entropy that crosses the event 
horizon.  This is given by the product of the comoving entropy density
and the largest comoving volume enclosed by the event horizon, which
occurs whenever the event horizon comes into existence.  
For concreteness, let us consider a spatially flat model for which the 
event horizon of an observer at $r=0$ meets the initial singularity at 
$r=\eta_{max}$, as is evident from figure~\ref{pen_normal}a.
 
The maximal conformal time given in equation (\ref{eulerb}) gives
\begin{equation}
\tilde{V}_{max} \sim \Lambda^{-n/2}
\end{equation}
for the maximal comoving volume enclosed by the event horizon.
The total entropy that passes through the light-sheet is thus
\begin{equation}
S = \tilde{s} \, \tilde{V}_{max} 
\sim G^{-\frac{n}{n+1}} \Lambda^{-\frac{n(n-1)}{n(n+1)}} .
\end{equation}
The ratio between this entropy and the horizon area at late times in 
Planck units is given by 
\begin{equation}
\frac{S}{{\cal A}_{dS} /G} 
\sim G^{\frac{1}{n+1}} \Lambda^{\frac{1}{2}\left(
    \frac{n-1}{n+1}\right)} .
\end{equation}
So we see that this ratio actually vanishes in the limit of small
$\Lambda$ and the covariant entropy bound is far from being saturated,
let alone violated, in this limit.

The story is different when we consider spatially flat
models with an equation of state such that 
$\beta \ge 1$ in (\ref{zerokappa}).
Models that exhibit this behavior include the $\kappa = 0$ solution 
for dust in 2+1 dimensions, which has $\beta =1$, and the negative
pressure models in section~\ref{negpress}, which have $\beta =2$.
The relevant Penrose diagram is now the one in
figure~\ref{pen_normal}b rather than figure~\ref{pen_normal}a. The
covariant entropy bound is violated because an infinite amount of
entropy will cross the event horizon of any comoving observer while
the horizon area remains bounded by the area of the corresponding de~Sitter
horizon. This conclusion could be avoided if the expansion of the past
directed null rays along the event horizon were to become 
positive at some time in the past, in which case the light-sheet would
terminate and the total entropy passing through it would be finite.
This cannot happen, however, as long as the area of the event horizon
increases with time.  A simple calculation for the $\kappa = 0$
negative pressure models, for example, shows that
\begin{equation}
{\cal A}_{eh}(\tau) = (1 - e^{-\tau}) {\cal A}_{dS} ,
\end{equation}
which clearly grows monotonically with $\tau$. The Penrose diagram in 
figure~\ref{pen_normal}b also applies to hyperbolic models with
$\kappa < 0$ and $\beta \ge 1$ and the above argument may be adapted
to arrive at a violation of the covariant entropy bound in these
models as well.

How seriously are we to take this violation of the covariant 
entropy bound?  The matter considered here has zero or negative 
pressure but does nevertheless obey the dominant energy condition.  
The corresponding cosmological models are perfectly good classical
solutions of Einstein gravity but a proof of the covariant entropy
bound has been given in classical gravity, under certain assumptions
about properties of the entropy fluid in spacetime \cite{fmw}.  At
first sight, our results appear to contradict that proof but a closer
look reveals that it is not so.  The source of the conflict with the
covariant entropy bound lies in the failure of conformal time to 
converge at the initial singularity in these models.  As a result the 
expansion starts off so slowly that an infinite comoving volume is 
visible in the causal past of any observer at finite $\tau >0$.
The geometry is singular at $\tau=0$ and a straightforward calculation
reveals that at very early times these big-bang solutions fail to meet 
the assumptions made in the proof of the covariant entropy bound
in \cite{fmw}.  Such early times are, however, beyond the reach of 
classical gravity and should not be included in a classical cosmological 
model \cite{kallin}.  By adapting arguments made in \cite{fissus} 
and \cite{kallin} to models with postitive $\Lambda$, one 
can show that if the covariant entropy bound is assumed to be satisfied 
at the Planck time it will not be violated at later times.  We therefore 
conclude that there is no conflict with the covariant entropy bound after 
all in any of our cosmological models but that in some cases the bound 
requires quantum effects to modify the early causal structure from that 
of classical big-bang solutions, changing Penrose diagrams of the form
shown in figure~\ref{pen_normal}b to diagrams as shown in 
figure~\ref{pen_normal}a.

\acknowledgments

We thank G.~Bj\"ornsson, R.~Bousso, U.~Danielsson, E.H.~Gudmundsson, 
D.~Marolf, and P.J.E.~Peebles for useful discussions.  This work was 
supported in part by grants from the Icelandic Research Council, The 
University of Iceland Research Fund, and The Icelandic Research Fund for 
Graduate Students.  K.R.K. would like to thank the Department of Physics at 
UC Santa Barbara for hospitality during this work.

\end{document}